\documentclass{article}
\usepackage[T1]{fontenc} 
\usepackage[utf8]{inputenc} 
\usepackage{waspaa19,amsmath,graphicx,times}
\usepackage{booktabs}
\usepackage{graphicx}
\usepackage{siunitx}
\usepackage{xcolor}

\usepackage[hidelinks]{hyperref}

\title{High-level control of drum track generation using learned patterns of rhythmic interaction\vspace*{-.3cm}}

\name{Stefan Lattner$^1$, Maarten Grachten$^2$ \vspace{-.3cm}}

\address{$^1$Sony Computer Science Laboratories (CSL), Paris, France\hspace{1em}$^2$ Contractor for Sony CSL, Paris, France}

\begin{document}
\maketitle
\begin{sloppy}

\begin{abstract}
  Spurred by the potential of deep learning, computational music generation has gained renewed academic interest.
  A crucial issue in music generation is that of user control, especially in scenarios where the music generation process is conditioned on existing musical material.
  Here we propose a model for conditional kick drum track generation that takes existing musical material as input, in addition to a low-dimensional code that encodes the desired relation between the existing material and the new material to be generated.
  These relational codes are learned in an unsupervised manner from a music dataset.
  We show that codes can be sampled to create a variety of musically plausible kick drum tracks and that the model can be used to transfer kick drum patterns from one song to another.
  Lastly, we demonstrate that the learned codes are largely invariant to tempo and time-shift.
\end{abstract}
\section{Introduction}\label{sec:introduction}
\vskip-1ex
A crucial issue in music generation is that of user control.
Especially for problems where musical material is to be generated conditioned on existing musical material (so-called \emph{conditional generation}), it is not desirable for a system to produce its output deterministically. 
Typically there are multiple valid ways to complement existing material with new material, and a music generation system should reflect that degree of freedom,
either by modeling it as a predictive distribution from which samples can be drawn and evaluated by the user, or by letting the generated material depend on some form of user input in addition to the existing material.
An intuitive way to address this requirement is to learn a latent space, for example by means of a variational autoencoder (VAE). This approach has been successfully applied to music generation~\cite{roberts18:_hierar_laten_vector_model_learn, simon2018learning}, and allows for both generation and manipulation of musical material by sampling from the latent prior, manual exploration of the latent space, or some form of local neighborhood search or interpolation.

In this paper we also take a latent space learning approach to address the issue of control over music generation.
More specifically, we propose a model architecture to learn a latent space that encodes rhythmic interactions of the kick drum vs. bass and snare patterns.
The architecture is a convolutional variant of a \emph{Gated Autoencoder} (GAE, see Section~\ref{sec:method}).
This architecture can be thought of as a feed-forward neural network where the weights are modulated by learned \emph{mapping codes}~\cite{DBLP:conf/iccv/Memisevic11}.
Each mapping code captures local relations between kick vs bass and snare inputs, such that an entire track is associated to a sequence of mapping codes.

Since we want mapping codes to capture rhythmic patterns rather than just the instantaneous presence or absence of onsets in the tracks, during training we enforce invariance of mapping codes to (moderate) time shifts and tempo changes in the inputs.
The resulting mapping codes remain largely constant throughout sections with a stable rhythm.
This provides high-level control over the generated material in the sense that different kick drum patterns for some section can be realized simply by selecting a different mapping code (either by sampling or by inferring them from another section or song), and applying it throughout the section.

To our knowledge this is a novel approach to music generation. It reconciles the notion of user control with the presence of conditioning material in a musically meaningful way: rather than controlling the characteristics of the generated material directly, it offers control over how the generated material \emph{relates} to the conditioning material. 

Apart from quantitative experiments to show the basic validity of our approach, we validate our model by way of a set of sound examples and visualized outputs. We focus on three scenarios specifically.
Firstly we demonstrate the ability to create a variety of plausible kick drum tracks for a given snare and bass track pair by sampling from a standard multivariate normal distribution in the mapping space.
Secondly, we test the possibility of \emph{style-transfer}, by applying rhythmic interaction patterns inferred from one song to induce similar patterns in other songs.
Finally, we show that the semantics of the mapping space is invariant under changes in tempo.

In continuation we present related work (Section~\ref{sec:related_work}), describe the proposed model architecture and data representations (Section~\ref{sec:method}), and validate the approach (Section~\ref{sec:experiments}).
Section~\ref{sec:conclusions} provides concluding remarks and future work.
\vspace{-1.2ex}
\section{Related work}\label{sec:related_work}
\vspace{-1.2ex}
In addition to the VAE-based methods for control over music generation processes mentioned above, a number of other studies have applied deep learning methods to address the problem of music generation in general, as reviewed in~\cite{DBLP:journals/corr/abs-1709-01620}.
Drum track generation has been tackled using recurrent architectures \cite{makris2018conditional, DBLP:conf/eann/MakrisKKK17}, Restricted Boltzmann Machines \cite{DBLP:conf/nime/VoglK17}, and Generative Adversarial Networks (GANs) \cite{dong2018musegan}.
Approaches to \emph{control} the generation process may rely on sampling from some latent representation of the material to be generated \cite{roberts18:_hierar_laten_vector_model_learn,simon2018learning}, possibly in an incremental fashion~\cite{hadjeres2017deepbach}, or conditioning on user-provided information (such as a style label~\cite{mao2018deepj}, unary~\cite{hadjeres2018anticipation}, or structural~\cite{lattnergeneration} constraints). 
\cite{DBLP:conf/icassp/GrinsteinDOP18} demonstrates style transfer for audio.
GANs are used in~\cite{dong2018musegan,liu18:_lead_sheet_gener_arran_condit}, where the output of the generation process is determined by providing some (time-varying) noise, in combination with conditioning on existing material.
Similar to our study, \cite{lattner2018predictive} uses a GAE to model \emph{relations} between musical material in an autoregressive prediction task.
To our knowledge this is the first use of GAEs for conditional music generation.

\vspace{-1.5ex}
\section{Method}\label{sec:method}
\vskip-1.8ex
A schematic overview of the proposed model architecture is shown in Figure~\ref{fig:model}.
For time series modeling, we adapt the common dense GAE architecture to 1D convolution in time, yielding a Convolutional Gated Autoencoder (CGAE).
We aim to model the rhythmic interactions between input signals $\mathbf{x} \in \mathcal{R}^{M \times T}$ and a target signal $\mathbf{y} \in \mathcal{R}^{1 \times T}$.
More precisely, $\mathbf{x}$ represents $M$ 1D signals of length $T$ indicating onset functions of instrument tracks and beat- and downbeat information of a song, while $\mathbf{y}$ represents the onset function of a target instrument.
Then the rhythmic interactions (henceforth referred to as \emph{mappings}) between $\mathbf{x}$ and $\mathbf{y}$ are defined as
\vspace{-0.8ex}
\begin{equation}\label{eq:gamap}
\mathbf{m} = \mathbf{W}*(\mathbf{U}*\mathbf{x} \cdot \mathbf{V}*\mathbf{y}),
\end{equation}
where $\mathbf{m} \in \mathcal{R}^{Q \times T}$, and $\mathbf{U} \in \mathcal{R}^{K \times M \times R}$, $\mathbf{V} \in \mathcal{R}^{K \times 1 \times R}$ represent respectively $K$ convolution kernels for $M$ input maps and kernel size $R$, and $\mathbf{W} \in \mathcal{R}^{Q \times K \times 1}$ represents $Q$ convolution kernels for $K$ input maps and kernel size $1$.
Furthermore, $*$ is the convolution operator and $\cdot$ is the Hadamard product.
For brevity the notation above assumes a CGAE architecture with only one mapping layer and one layer for input and target.
In practice we use several convolutional layers, as described in Section \ref{sec:train_details}.

\begin{figure}
\begin{center}
\includegraphics[trim=0cm 0cm 2cm 0cm, clip, width=.65\linewidth]{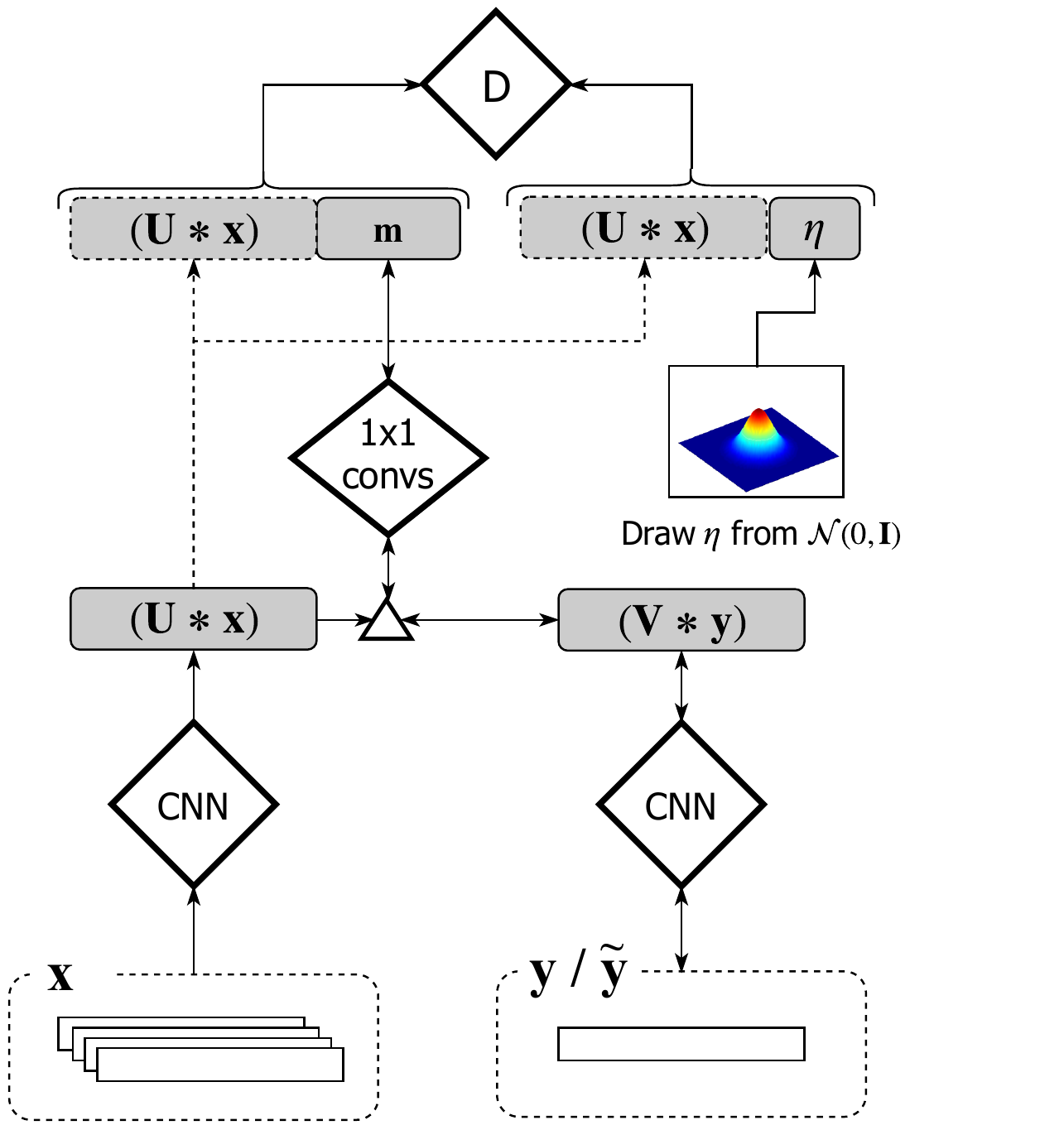} \\
\end{center}
\vskip-3ex
\caption{The proposed model architecture.}
\vskip-2ex
\label{fig:model}
\end{figure}

Given the rhythmic interactions $\mathbf{m}$ and the rhythmic context $\mathbf{x}$, the target onset function is reconstructed as
\vspace{-0.8ex}
\begin{equation}\label{eq:recon}
\mathbf{\tilde{y}} = \mathbf{V}^\top *(\mathbf{U}*\mathbf{x} \cdot \mathbf{W}^\top *\mathbf{m}),
\end{equation}
where the transposed kernels $\mathbf{V}^\top$ and $\mathbf{W}^\top$ result in a deconvolution.
The model parameters are trained by minimizing the mean squared error $\mathcal{L}_{\text{mse}}(\mathbf{y}, \mathbf{\tilde{y}})$ between the target signal $\mathbf{y}$ and its reconstruction $\mathbf{\tilde{y}}$.

In order to draw samples from the model, we want to impose a Gaussian prior over $\mathbf{m}$ resulting in $p(\mathbf{m}) = \mathcal{N}(0,\mathbf{I})$.
Additionally, $\mathbf{m}$ should apply to any input $\mathbf{x}$, and should therefore not contain any information about the content of $\mathbf{x}$.
These conditions are imposed using adversarial training \cite{DBLP:journals/corr/MakhzaniSJG15}:
A discriminator $D(\cdot)$ estimates whether its input is drawn from a Gaussian distribution and contains no information about $\mathbf{x}$.
To that end, we concatenate $(\mathbf{U}\!*\!\mathbf{x})$ with either actual mappings $\mathbf{m}$ or noise drawn from an independent Gaussian distribution $\boldsymbol{\eta} \sim \mathcal{N}(0,\mathbf{I}), \boldsymbol{\eta} \in \mathcal{R}^{Q \times T}$.
This results in $D(\mathbf{m}, (\mathbf{U}*\mathbf{x}))$ and $D(\boldsymbol{\eta}, (\mathbf{U}*\mathbf{x}))$.
In adversarial training, the discriminator $D(\cdot)$ learns to distinguish between the input containing $\mathbf{m}$ and the input containing $\boldsymbol{\eta}$.
If there is mutual information between $(\mathbf{U}*\mathbf{x})$ and $\mathbf{m}$ the discriminator can exploit this for its classification task.
This causes the encoding pathways to remove any information about $\mathbf{x}$ from $\mathbf{m}$.
Also, we obtain $\mathbf{m} \sim \mathcal{N}(0,\mathbf{I})$.
Accordingly, the discriminator is trained to minimize the loss
\vskip-2ex
\begin{equation}\label{eq:loss_discrim}
\mathcal{L}_{\text{advers}} = \frac{1}{T} \sum_t{D(\mathbf{m}, (\mathbf{U}*\mathbf{x}))_t - D(\boldsymbol{\eta}, (\mathbf{U}*\mathbf{x}))_t},
\vspace{-.5em}
\end{equation}
with $D(\cdot)_t$ being the output of the discriminator at time $t$.

To make the mappings more constant over time, an additional loss penalizes differences of successive mappings $\mathcal{L}_{\text{const}} = \frac{1}{T}\sum_t{(\mathbf{m}_t - \mathbf{m}_{t+1})^2},\mathbf{m}_t \in \mathcal{R}^{Q}$.
A further loss that constrains each map $\mathbf{m}_q \in \mathcal{R}^{T}$ to have zero mean and unit variance over time and instances in a batch considerably improves the learning of the CGAE:
\begin{equation}
\mathcal{L}_{\text{std}} = \frac{1}{Q} \sum_q^Q \bigg[\Big(\frac{1}{N} \sum_i^N{(m_{q,i} - \mu_q)^2}\Big) - 1 \bigg]^2 + \mu_q^2,
\vspace{-.5em}
\end{equation}
where $m_{q,i}$ are the observations of convolutional map $\mathbf{m}_q$ over all time steps and instances in a batch, and $\mu_q$ is the mean of $m_{q,i}$.
Optimization is performed in two steps per mini-batch.
First, the discriminator is trained to minimize $\mathcal{L}_{\text{advers}}$, then the CGAE is trained to minimize $\mathcal{L}_{\text{mse}}(\mathbf{y}, \mathbf{\tilde{y}}) + \mathcal{L}_{\text{const}} + \mathcal{L}_{\text{std}} - \mathcal{L}_{\text{advers}}$.

\vspace{-.5em}
\subsection{Architecture and training details}\label{sec:train_details}
\vspace{-.5em}
As mentioned above the weight matrices $\mathbf{W}, \mathbf{U}$ and $\mathbf{V}$ in Eqs.~\ref{eq:gamap} and~\ref{eq:recon} act as placeholders for several convolutional layers.
For $\mathbf{U}$ and $\mathbf{V}$, $8$ convolutional layers are defined, with $\{32, 32, 64, 64, 64, 128, 128, 256\}$ output units, kernel size $2$, and dilations which double for each layer (i.e., $1,2,4,8,\dots$).
The first $5$ layers keep the 4 inputs (onset strength snare, onset strength bass, beats, downbeats) separated (i.e., their units are separated in $4$ groups, where each group only pools over $1/4$th of the input maps), and the information is combined only in the two top-most layers.
$\mathbf{W}$ is a placeholder for $6$ layers, with sizes $\{128, 128, 64, 32, 32, 16\}$, and kernel size $1$.
The discriminator $D(\cdot)$ consists of $5$ layers with $\{256, 128, 64, 64, 1\}$ maps and kernel size $1$.
All stacks described above have no non-linearity in the output and SELU non-linearity \cite{DBLP:conf/nips/KlambauerUMH17} between layers (also for the de-convolution passes).

The model is trained for $2500$ epochs with batch size $100$, using $50\%$ dropout on the inputs $\mathbf{x}$.
During training, a data augmentation based regularization method is used to make the mappings invariant to time shift and tempo change.
To that end we define a transformation function $\psi_\theta(\mathbf{z})$ that shifts and scales a signal $\mathbf{z}$ in the time dimension with random shifts between $-150$ and $+150$ time steps ($\pm1.75s$) and random scale factors between $0.8$ and $1.2$.
Training is then performed as follows.
First, the mappings $\mathbf{m}$ are inferred according to Eq.~\ref{eq:gamap}.
Then, the input signals are modified using $\psi_\theta(\cdot)$ resulting in an altered Eq.~\ref{eq:recon}:
\begin{equation}\label{eq:recon_aug}
\mathbf{\tilde{y}_{\psi_\theta}} = \mathbf{V}^\top *(\mathbf{U}* \psi_\theta(\mathbf{x}) \cdot \mathbf{W}^\top *\mathbf{m}).
\end{equation}
\vskip-1ex
Finally, the mean squared error between the such obtained reconstruction and the \emph{transformed} target is minimized as $\mathcal{L}_{\text{mse}}(\psi_\theta(\mathbf{y}), \mathbf{\tilde{y}_{\psi_\theta}})$.
\noindent This approach was first proposed in~\cite{lattner2018learning}.
Due to the gating mechanism $\cdot$ (activating only pathways with appropriate tempo and time-shift), a CGAE is particularly suited for learning such invariances.
By imposing time-shift invariance, we assume that rhythmic interaction patterns (and the respective mappings) in the training data are locally constant.
Even if this method introduces some error at positions where rhythmic patterns change, most of the time the assumption of locally constant rhythm is valid.
\vspace{-1.5em}

\subsection{Data representation}
\vskip-1ex
The training/validation sets consist of $665 / 193$ pop/rock/ electro songs where the rhythm instruments bass, kick and snare are available as separate 44.1kHz audio tracks.
The context signals $\mathbf{x}$ consist of two 1D input maps for beat and downbeat probabilities, and two 1D input maps for the onset functions of Snare and Bass.
The target signal $\mathbf{y}$ consists of a 1D onset function of the Kick drum.
The onset functions are extracted using the \emph{ComplexDomainOnsetDetection} feature of the publicly available Yaafe library\footnote{\url{http://yaafe.sourceforge.net}} with a block size of $1024$, a step size of $512$, and a Hann window function.
For the downbeat functions we use the downbeat estimation RNN of the madmom library\footnote{\url{https://github.com/CPJKU/madmom}}.
Input signals are individually standardized to zero mean and unit variance over each song.
\vspace{-.5em}

\subsection{Rendering Audio}\label{sec:rendering}
\vskip-1ex
We create an actual kick drum track from an onset strength curve $\mathbf{y}$ using salient peak picking.
First, we remove less salient peaks from $\mathbf{y}$ by zero-phase filtering with a low-pass Butterworth filter of order two and a critical frequency of half the Nyquist frequency. 
The local maxima of the smoothed curve are then thresholded, discarding all maxima below a certain proportion (see below) of the maximum of the standardized onset strength curve.
The remaining peaks are selected as onset positions.
Finally, we render an audio file by placing a ``one-shot'' drum sample on all remaining peaks after thresholding.
We introduce dynamics by choosing the volume of the sample from $70 \%$ for peaks at the threshold to $100 \%$ for peaks with the maximum value.
For the qualitative experiments in the following section, we manually choose the threshold between $15 \%$ and $50 \%$.
For the quantitative results in Table~\ref{tab:fscores} we fix the threshold at $25 \%$, but values of $20 \%$ and $30 \%$ yield similar figures.
\vspace{-1.5ex}
\section{Experiments}\label{sec:experiments}
\vspace{-1.5ex}
For the  qualitative experiments we use four songs, \emph{Gipsy Love}, \emph{Orgs Waltz}, \emph{Miss You} and \emph{Drehscheibe}, produced by the first author.
We encourage the reader to listen to the results on the accompanying web page\footnote{\url{https://sites.google.com/view/drum-generation}}.
Three scenarios are chosen to show the effectiveness of the proposed approach:


\textbf{Conditional Generation of Drum Patterns}
\ To generate a kick drum track, we sample only one mapping code $\mathbf{m}_t$ (from a 16-dimensional standard Gaussian), repeat it across the time dimension, and reconstruct $\mathbf{y}$ given the resulting $\mathbf{m}$, as well as $\mathbf{x}$.
Subsequently, we render $20$ audio files as described in Section \ref{sec:rendering} and pick those $10$ which together constitute the most varied set.
Figure \ref{fig:samples_kick} shows some results of the generation task -- randomly generated kick drum tracks conditioned on the song \emph{Drehscheibe} (the sound examples are available online).
It is clear from these screenshots that the model generates a wide variety of different rhythmic patterns which adapt to the local context, even though the sampled mapping code is constant (repeated) over time.

\begin{figure}[t]
\begin{center}
\includegraphics[trim=10.8cm 1.7cm 0cm 1.8cm, clip, width=1.\linewidth]{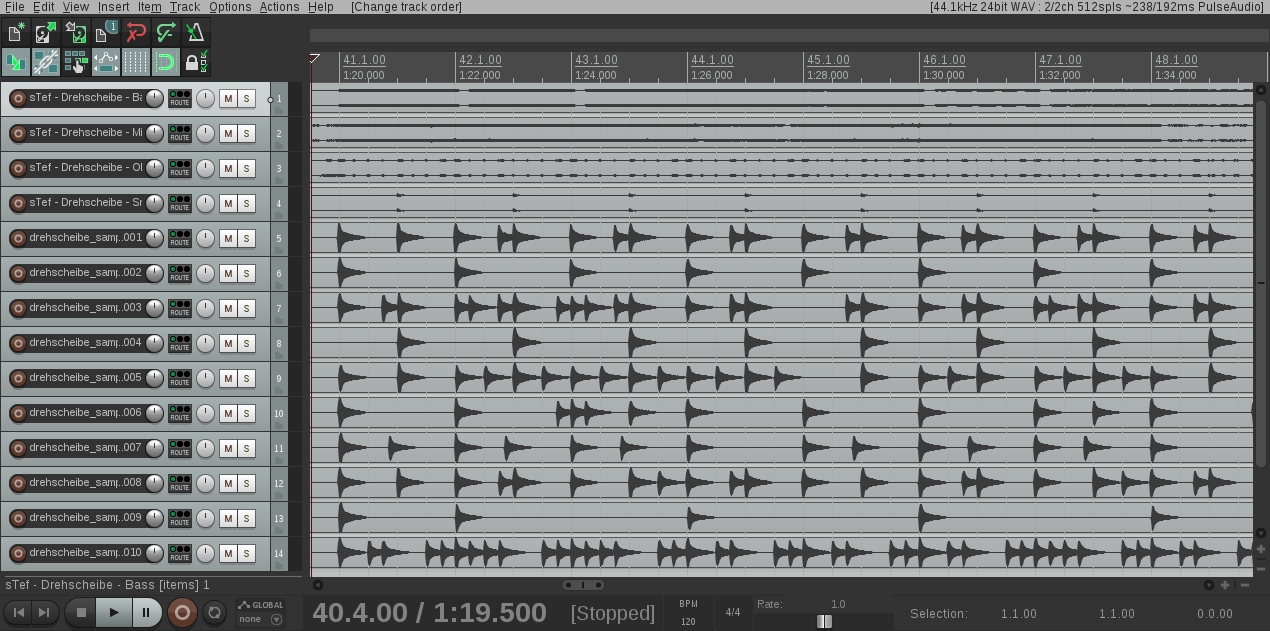} \\
\end{center}
\vspace{-.5cm}
\caption{Conditionally sampled kick drum tracks for the song \emph{Drehscheibe}. Each track is the result of a sampled ``transfer function'' $\mathbf{m}_t$ which is held constant over time.}
\label{fig:samples_kick}
\end{figure}

\textbf{Style Transfer}
\ First, for a given song, we infer $\mathbf{m}$ from $\mathbf{x}$ and $\mathbf{y}$.
Second, k-means clustering is performed over all $\mathbf{m}_t$, using the Davis-Bouldin score \cite{DBLP:journals/pami/DaviesB79} for determining the optimal number of clusters (typically yielding an optimal k between 5 and 8).
Then we use the cluster center of the largest cluster found as the mapping code, again repeat it over time and use it for another song onto which the style should be transferred.
Again the results are available on the accompanying web page (see above).

\textbf{Tempo-invariance}\ ~We use the WSOLA time stretching algorithm \cite{verhelst93:_wsola} as implemented in \emph{sox}, 
to create four time stretched versions of each song, at $80\%$, $90\%$, $110\%$ and $120\%$ of the original tempo, respectively.
Then, for a given song in \emph{original tempo}, we determine a prototypical mapping code with the k-means clustering method described above.
We repeat that code througout the \emph{time-stretched} versions of the song and reconstruct $\mathbf{y}$ given $\mathbf{m}$ and $\mathbf{x}$.
Figure \ref{fig:tempo_invar} shows generated kick drum tracks in the five different tempos (four time-stretched versions plus the original tempo).
It is clear from these screenshots that the drum pattern adjusts to the different tempos and does not change its style.

\begin{figure}[t]
\centering
\includegraphics[width=.75\linewidth]{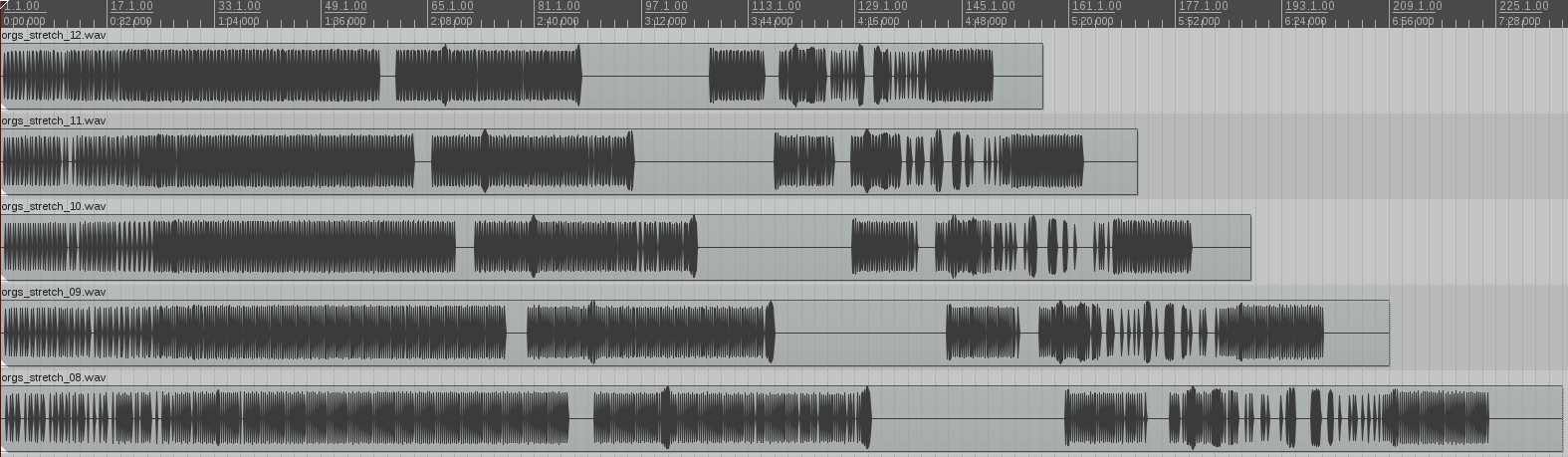} \\
\vspace{1mm}
\includegraphics[width=.75\linewidth]{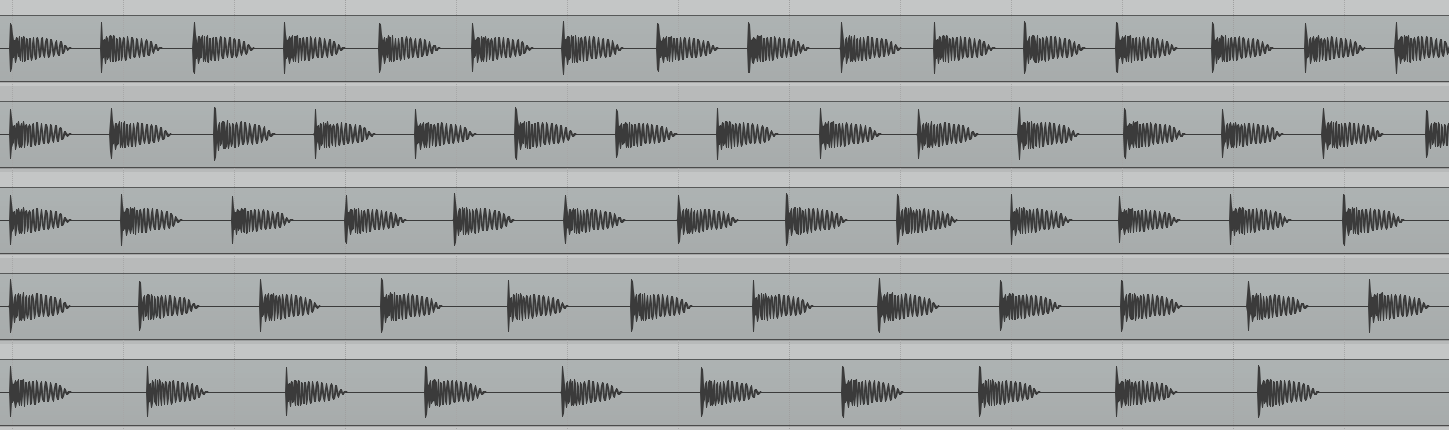} \\
\vspace{-.1cm}
\caption{Generated drum tracks for different tempos ($80\%$, $90\%$, $100\%$, $110\%$ and $120\%$ of the original tempo) for the song \emph{Orgs Waltz}, using the mapping code of the original tempo. Top: the overall song; Bottom: close-up of the first onsets, manually aligned for the purpose of visualization.}
\label{fig:tempo_invar}
\vskip-2ex
\end{figure}


Although it is not obvious how to evaluate the output of the model other than by listening, we can check the validity of basic assumptions about the behavior of the model.
One assumption is that the \emph{ground truth} mappings for a song---as defined in Eq.~(\ref{eq:gamap})---allow us to reconstruct the drum track relatively faithfully (Eq.~(\ref{eq:recon})).
To test the degree to which reconstruction may be sacrificed to satisfy other constraints (e.g., the adversarial loss) we compute the accuracy of the reconstruction.
Given onset strength curves $\mathbf{y}$ and $\mathbf{\tilde{y}}$ we determine the onset positions as described in Section~\ref{sec:rendering}, and compute the precision, recall, and F-score using a 50 ms tolerance window, following MIREX onset detection evaluation criteria~\cite{mirexonsetdetection2018}.

Table~\ref{tab:fscores} (upper half) lists the results for the training and validation sets and shows that the mappings are specific enough to largely reconstruct the target onsets correctly.
The reconstructions are not perfect, likely due to the model's invariance and the adversarial loss on the mappings.
Note also that the accuracy for the validation set is similar to that for the training set, implying that no overfitting has occurred.
The dominance of precision over recall is likely due to the typical ``conservative'' behavior of GAEs \cite{ICML2012Memisevic_105}.

\begin{table}
  \footnotesize
  \sisetup{
    round-mode = places,
    round-precision = 3,
    table-format=1.3
  }
  \centering
  \renewcommand{\arraystretch}{1.1}
  \begin{tabular}{cSSS}
    \toprule
    & {Precision} & {Recall} & {F-Score} \\ \hline
    \emph{Ground truth} & & & \\ \hline
    Training     &  0.945684210526 & 0.81222556391 & 0.864676691729 \\
    Validation   &  0.942849740933 & 0.816476683938 & 0.867305699482 \\ \hline
    \emph{Style transfer} & & & \\ \hline
    Training & 0.773804511278 & 0.695669172932 & 0.712421052632 \\
    Validation & 0.781450777202 & 0.707409326425 & 0.723212435233 \\
    \bottomrule
  \end{tabular}
  \caption{Average precision, recall, and F-score for onset reconstruction using ground truth and style transfer mappings.}
  \vspace{-.3cm}
  \label{tab:fscores}
\end{table}

Furthermore we test the validity of the heuristic of taking the largest cluster centroid as a constant mapping vector over time for style transfer (assuming time-invariance). 
To do so, we apply this heuristic to transfer the style of a song to itself.
That is, to reconstruct the kick drum track we use the largest mode of the song in the mapping space as a constant through time, rather than the ground truth mapping---a trajectory through the mapping space.
Unsurprisingly, this approximation affects the reconstruction of the original kick drum track negatively, but the F-scores of over $0.7$ still shows that a substantial part of the tracks is reconstructed correctly.


\vspace{-.4em}
\section{Conclusions and future work}\label{sec:conclusions}
\vskip-1ex
We have presented a model for the conditional generation of kick drums tracks given snare and bass tracks in pop/rock/electro music.
The model was trained on a dataset of multi-track recordings, using a custom objective function to capture the relationship between onset patterns in the tracks of the same song in mapping codes.
We have shown that the mapping codes are largely tempo and time-invariant and that musically plausible kick drum tracks can be generated given a snare and bass track either by sampling a mapping code or through style transfer, by inferring the mapping code from another song.

Importantly, two basic aspects of the chosen approach have been shown to be valid.
Firstly, the ground-truth mapping codes are able to faithfully reconstruct the original kick drum track.
Secondly, the style transfer heuristic of applying a constant mapping code through time was shown to be largely valid, by comparing the original kick drum track of a song to the result of applying the style of a song to itself.

Although the current work is limited in the sense that the model has only been demonstrated for kick drum track generation, we believe this approach is applicable to other content.
We are currently applying the same approach to snare drum generation and $\mathit{f}0$-generation for bass tracks.

\section{Acknowledgements}\label{sec:acknowledge}
We thank Cyran Aouameur for his valuable support, as well as Adonis Storr, Tegan Koster, Stefan Weißenberger and Clemens Riedl for their contribution in producing the example tracks.

\bibliographystyle{IEEEtran}
\bibliography{mg,sl}
\end{sloppy}
\end{document}